# An Insight Analysis of Nano sized powder of Jackfruit Seed

T. Theivasanthi* and M. Alagar

Centre for Research and PG.Department of Physics, Ayya Nadar Janaki Ammal College, Sivakasi-626124, Tamilnadu, India.
*Corresponding author. *E-mail*: theivasanthi@pacrpoly.org

___________________________________________________________________________

*Abstract: The preparation of biodegradable nanomaterials by blending starch nanocrystals with various polymer matrices are the most active research. This work reports aspect related to nano-sized particles of jackfruit seed. This approach is simple, faster, eco-friendly, cost effective and suitable for large scale production. X-Ray Diffraction studies analyze particles size, morphology, type of starch and degree of crystallinity. The particle size is found to be 12nm, specific surface area is 625 $m^2g^{-1}$, contains A-type starch and 32% degree of crystallinity. A morphology index (MI) is developed from FWHM of XRD data to understand interrelationship of particle size and specific surface area. MI range is from 0.50 to 0.74. It is correlated with the particle size and specific surface area. It is observed that MI has direct relationship with particle size and an inverse relationship with specific surface area.*

___________________________________________________________________________

## 1. Introduction

The *Artocarpus heterophyllus* Lam is a species of tree of the mulberry family (*Moraceae*) is commonly known as jackfruit. It is native to Western Ghats of India and Malaysia. It produces heavier yield than any other tree and bear the largest known edible fruit (up to 35kg). Many parts of this plant, including the bark, roots, leaves, fruits and seeds have medicinal properties.

The sweet yellow sheaths around the seeds are about 3-5 mm thick and have a milder and less juicy. Seeds are separated horny endocarpus enclosed by sub-gelatinous exocarpus (1mm thick) a thin whitish membrane. They are oval, oblong or oblong ellipsoid or rounded shape, light brown colour in nature, 2-3 cm (0.8-1.2 inch) in length and 1-1.5 cm (0.4-0.6 inch) in diameter. Up to 500 seeds can be found in each fruit. They are recalcitrant and can be stored up to a month in cool, humid conditions [1].

Jackfruit seeds are nutritious, rich in potassium, fat, carbohydrates and minerals. Manganese and magnesium elements have also been detected in seed powder [2]. Seeds contain two lectins namely jacalin and artocarpin. Jacalin has been seen to inhibit the herpes simplex virus type 2 and has proved to be useful for the evaluation of the immune status of patients infected with human immunodeficiency virus 1 (HIV1). It is also used for the isolation of human plasma glycoproteins, the investigation of IgA nephropathy, the analysis of 0-linked glycoproteins and the detection of tumours [3].

Starch-based materials originally attracted a great deal of interest because of their low cost, real biodegradability, and renewable origins. Thermoplastic starch (TPS) is one of such materials obtained after disruption and plasticization of starch by heating in presence of water or other plasticizers such as glycerol. Thermoplastic starches have being also successfully blended with inorganic material like clay, other suitable polymers like natural rubber [4].

Starch is a biocompatible polymer. It is useful in making hybrid organic-inorganic materials, hybrid Composites, starch /clay- composite and Polymer/Clay nano-composites. Reduced defects, increased surface area, percolation, interphase volume, polymer morphology are the concepts of nanocomposites. As the size of a particle is reduced, the number of defects per particle is also reduced and mechanical properties rise proportionately.

Starch (carbohydrate) is occupying the major proportion of Jackfruit seeds powder. Because of much starch, XRD pattern is mainly showing the properties of starch. A survey of literature indicates that not much work has been done on the jackfruit seed. It also appears from the limited published data that no complete XRD investigations on the nanosized particles of jackfruit seed has so far been carried out. For the first time, XRD investigations of the nanosized particles of jackfruit seed are recorded. This work is expected to throw some light on and help further research.

## 2. Experimental Details

Ripe jackfruit (Artocarpus heterophyllus) seeds were collected from the foot hills of Western Ghat hills at Rajapalayam, Tamilnadu, India (a place mostly free from industrial pollution) and used for this study. The seeds were cleaned and the white arils (seed coats) were peeled off. Seeds were sun dried for 7 days without remove the thin brown spermoderm covers the fleshy white cotyledons. For these experimental purposes, 100gm dried seeds were put in a mixer-grinder cum blender which having 550 watts, 17000 rpm rotating speed electrical motor. The seeds were grinded and crushed well and uniformly for 15 minutes with utmost precaution to avoid any contamination and made them as nano-sized powder. The powdered materials were packed in plastic pouches and stored in normal room temperature until use.

XRD analysis of the prepared sample of jackfruit seed nanoparticles was done using a X'pert PRO of PANalytical diffractometer, Cu-Kα X-rays of wavelength ($\lambda$)=1.54056 Å and data was taken for the 2θ range of 10° to 80° with a step of 0.02°. XRD analysis gave size and degree of crystallinity of the particles. Its structural characterizations were studied and results confirmed the nano sized powder of jackfruit seed.

## 3. Results and Discussions

### 3.1. X-Ray Diffraction Studies - Peak Indexing

The X-ray diffraction pattern of the jackfruit seed nanoparticles is shown in Fig.1. Indexing process of powder diffraction pattern is done and *Miller Indices* (h k l) to each peak is assigned in first step. The details are in Table.1. A number of Bragg reflections can be seen which correspond to the (111), (200), (211) (311) and (322) reflections. The diffraction peaks are broad which indicating that the crystallite size is very small. The size of the jackfruit seed nanoparticles estimated from Debye–Scherrer formula (Instrumental broadening) is 12nm.

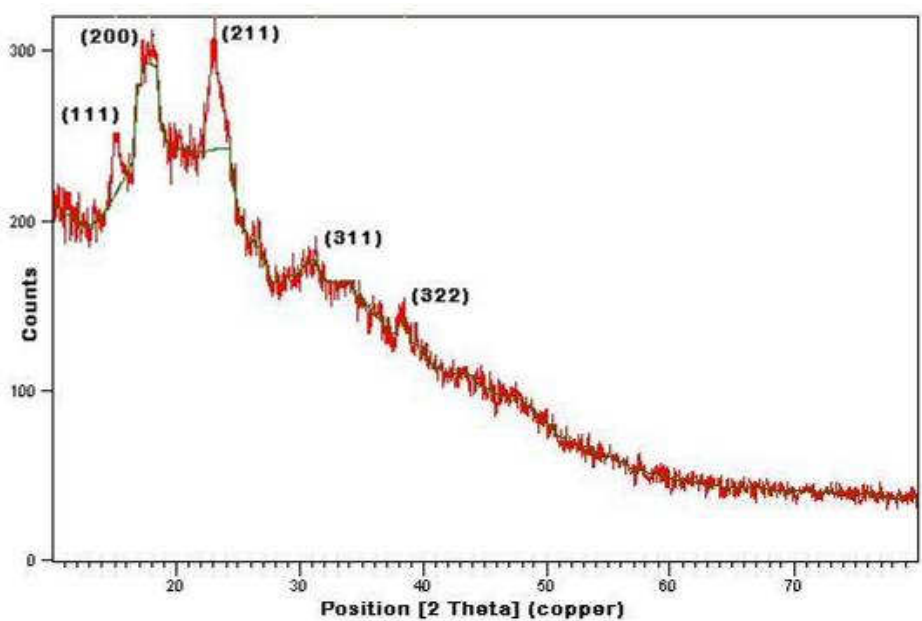

**Figure 1.** XRD showing Peak Indices & 2 θ Positions.

**Table 1.** Peak indexing from d – spacing.

| 2θ | d ( Å ) | 1000/d² | (1000/d²) / 10.6 | hkl |
|---|---|---|---|---|
| 15.1463 | 5.84966 | 29.224 | 3 | 111 |
| 17.7136 | 5.00722 | 39.835 | 4 | 200 |
| 23.0532 | 3.85811 | 67.132 | 6 | 211 |
| 31.2374 | 2.86345 | 121.361 | 11 | 311 |
| 38.3799 | 2.34541 | 181.787 | 17 | 322 |

## 3.2. XRD - Particle Size Calculation

From this study, considering the peak at degrees, average particle size has been estimated by using *Debye-Scherrer formula*.

$$D = \frac{0.9\lambda}{\beta \cos\theta} \quad \ldots\ldots\ldots\ldots\ldots\ldots\ldots\ldots\ldots\ldots\ldots\ldots\ldots (1)$$

Where 'λ' is wave length of X-Ray (0.1541 nm), 'β' is FWHM (full width at half maximum), 'θ' is the diffraction angle and 'D' is particle diameter size. The calculated particle size details are in Table.2. The value of d (the interplanar spacing between the atoms) is calculated using *Bragg's Law*.

$$2d\sin\theta = n\lambda \quad \ldots\ldots\ldots\ldots\ldots\ldots\ldots\ldots\ldots\ldots\ldots\ldots\ldots (2)$$

**Table 2.** The grain size of Jackfruit seed nanopowder.

| 2θ of peak (deg) | hkl | FWHM of peak (β) radians | Size of the partcle (D) nm | d-spacing nm |
|---|---|---|---|---|
| 15.1463 | (111) | 0.00930 | 15 | 0.584966 |
| 17.7136 | (200) | 0.00450 | 32 | 0.500722 |
| 23.0532 | (211) | 0.01167 | 12 | 0.385811 |
| 31.2374 | (311) | 0.00560 | 26 | 0.286345 |
| 38.3799 | (322) | 0.00405 | 37 | 0.234541 |

## 3.3. XRD - Instrumental Broadening

When particle size is less than 100 nm, appreciable broadening in x-ray diffraction lines will occur. Diffraction pattern will show broadening because of particle size and strain. The observed line broadening will be used to estimate the average size of the particles. The total broadening of the diffraction peak is due to the sample and the instrument. The sample broadening is described by

$$FW(S) \times cos\theta = \frac{K \times \lambda}{Size} + 4 \times Strain \times sin\theta \quad \ldots\ldots\ldots\ldots (3)$$

The total broadening $\beta_t$ is given by the equation

$$\beta_t^2 \approx \{\frac{0.9\lambda}{Dcos\theta}\}^2 + \{4\varepsilon tan\theta\}^2 + \beta_0^2 \quad \ldots\ldots\ldots\ldots (4)$$

$\varepsilon$ is strain and $\beta_0$ instrumental broadening. The average particle size D and the strain $\varepsilon$ of the experimentally observed broadening of several peaks will be computed simultaneously using *least squares method*. Instrumental Broadening is presented in Figure.2.

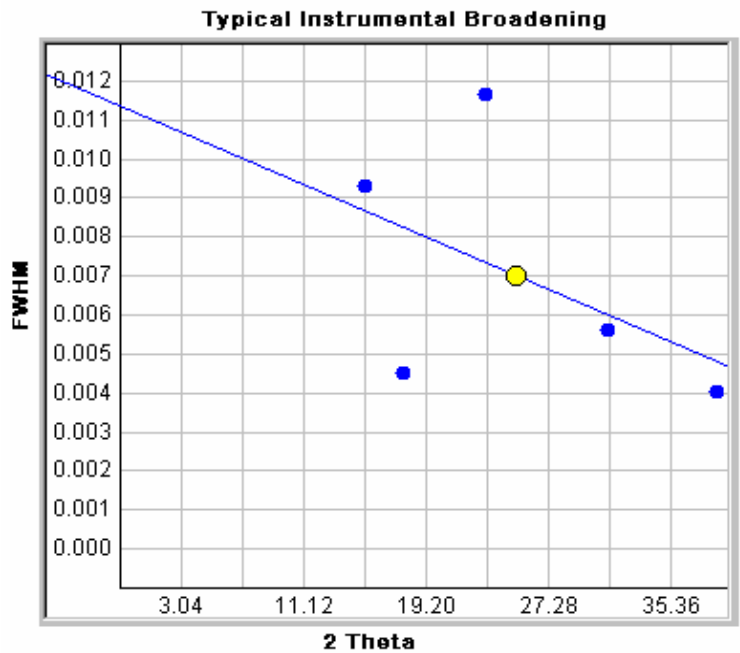

**Figure 2.** Typical Instrumental Broadening. y = -0.000166x + 0.0112.

Williamson and Hall proposed a method for deconvoluting size and strain broadening by looking at the peak width as a function of 2θ. Here, Williamson-Hall plot is plotted with sin θ on the x-axis and β cos θ on the y-axis (in radians). A linear fit is got for the data. From the linear fit, particle size and strain are extracted from y-intercept and slope respectively. The extracted particle size is 12 nm and strain is 0.005. Figure.3. shows Williamson Hall Plot.

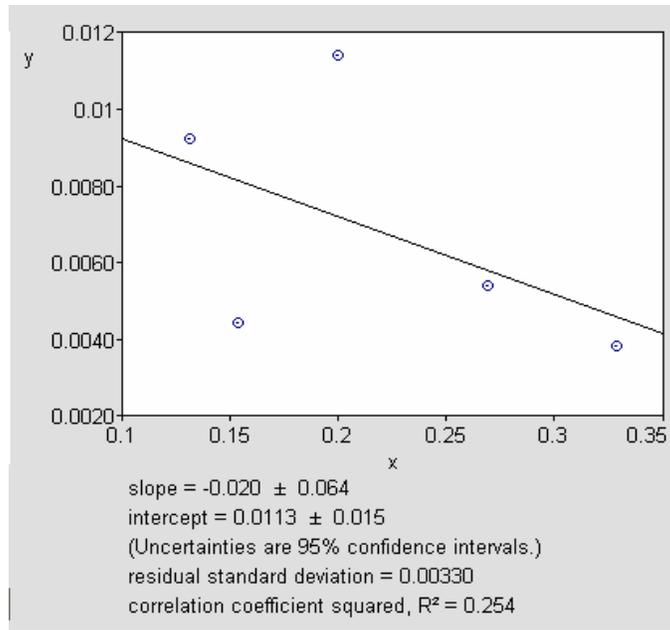

**Figure 3.** Williamson Hall Plot is indicating line broadening value due to the equipment.

Line broadening analysis is most accurate when the broadening due to particle size effects is at least twice the contribution due to instrumental broadening. The size range is calculated over which this technique will be most accurate. A rough upper limit is estimated for reasonable accuracy by looking at the particle size that would lead to broadening equal to the instrumental broadening. For example, for Monochromatic Lab X-ray (Cu $K\alpha$ FWHM ~ 0.05° at 20° 2θ), the accurate Size Range is < 90 nm (900 Å) and the rough Upper Limit is = < 180 nm (1800 Å).

**3.4. XRD - Starch Analysis**

Much of the information about starch granule crystalline properties has been acquired from Powder XRD studies. Starch can be classified into A, B and C forms [5]. A form starch is mainly present in cereal starches, such as maize starch and wheat starch. The XRD patterns of these starches give the stronger diffraction peaks at around 15, 17, 18 and 23°. The B form starch is usually available in tuber starch such as potato and this type of starch gives the strongest diffraction peak at 17° 2θ. There were also few small peaks at around 2θ values of 20, 22 and 24°. The C pattern starch is a mixture of both A and B types, such as smooth-seeded pea starch and various bean starches.

Because of much starch in jackfruit seed, XRD pattern is mainly showing the properties of starch. It shows diffraction peaks at around 15°, 17°, 23°, 31°and 38°. There is no peak in the range 5 °- 6 ° of 2θ like potato starch. The absence of peak in the range 5 °- 6 ° is characteristic of type-*A* starch [6]. Transition from A-type starch to B type pattern occurs at amylose content about 40% [7]. Earlier literature study also reveals amylose content of jackfruit seed starch is ~32% only [8]. These are indicating the presence of type-A starch in jackfruit seed.

**3.5. XRD- Degree of Crystallinity**

A starch granule is biosynthesized semi-crystalline granules containing densely packed polysaccharides and a small amount of water, being comprised of crystalline and amorphous domains. The inner structure of starch is that it is formed from two regions—crystalline and amorphous lamellae, which together form the crystalline and amorphous growth rings. When heated in excess water, starch granules undergo an ordered-disorder transition known as gelatinization. This phenomenon is associated with loss of crystallinity indicated by the disappearance of birefringence [9].

Starch is an important polysaccharide reserve in plants. It consists of two main components, amylose and amylopectin. It is a semi-crystalline polymer in which amylose forms the crystalline region and amylopectin forms the amorphous region [10]. The main component in the Jackfruit seed powder is starch which occupying approximately 78% content in the total biomass. The main crystalline peaks in the XRD pattern are attributed to the crystalline peaks of starch.

The jackfruit seed starch granules were apparently not susceptible to breakdown by thermal or mechanical sheer, indicating that the bonding forces within the granule remained strong even though the granules underwent gelatinization and swelled to a high degree 95 ˚C [11].

It is generally agreed that the peak breadth of a specific phase of material is directly proportional to the mean crystallite size of that material. Quantitatively speaking, sharper XRD peaks are typically indicative of high nanocrystalline nature and larger crystallite materials. From our XRD data, a peak broadening of the nanoparticles is noticed.

Nara and Komiya are calculating degree of crystallinity using Smadchrom software [12].

$$X_c = A_c/(A_c + A_a) \quad \ldots \ldots \ldots \ldots \ldots \ldots (5)$$

Where Xc = refers to the degree of crystallinity, Ac = refers to the crystallized area, A$a$ = refers to the amorphous area.

An empirical method of Segal for degree of crystallinity calculation is below [13].

$$CrI = 100 \frac{I_{002} - I_{Amorph}}{I_{002}} \quad \ldots \ldots \ldots \ldots (6)$$

Where CrI is the degree of crystallinity, $I_{002}$ is the maximum intensity of the (002) lattice diffraction and $I_{Amorph}$ is the intensity (311) diffraction at 31° 2θ degrees.

The maximum intensity of experimental jackfruit seed powder XRD is 102.85 for (002) lattice diffraction and intensity 69.16 for diffraction (311). Peak intensity details are enumerated in Table.3. CrI is calculated using the Segal et al. equation and as per calculation CrI is 32.75. This value is well in agreement with amylose content of starch of jackfruit seed powder 32% and amylose occupies the crystalline parts of starch.

Table 3. Intensity of XRD Peaks

| hkl | 111 | 200 | 211 | 311 | 322 |
|---|---|---|---|---|---|
| 2θ of peak (deg) | 15.1463 | 17.7136 | 23.0532 | 31.2374 | 38.3799 |
| Height (counts) | 36.29 | 102.85 | 63.50 | 69.16 | 43.00 |
| Relative Intensity (%) | 35.28 | 100.00 | 61.74 | 67.24 | 41.81 |

### 3.6. XRD - Specific Surface Area

Specific surface area (SSA) is a material property. It is a derived scientific value that can be used to determine the type and properties of a material. It has a particular importance in case of adsorption, heterogeneous catalysis and reactions on surfaces. SSA is the SA per mass.

$$SSA = \frac{SA_{part}}{Vpart * density} \quad \ldots \ldots \ldots \ldots \ldots \ldots (7)$$

Here SSA is Specific surface area, SApart is Surface Area of particle, Vpart is particle volume and density is jackfruit seed powder density [14].

$$S = 6 * 10^3 / D_p \rho \quad \quad \quad \quad \quad (8)$$

Where S is the specific surface area, Dp is the size of the particles, and ρ is the density of jackfruit seed powder [15]. Mathematically, SSA can be calculated using these formulas. Both of these formulas yield same result. The theoretical calculation of surface area, volume, SSA of jackfruit seed nanoparticles are presented in Table.4. The value of density is 0.80g/cm$^3$ which noted from the studies of F.C.K.Ocloo et al. [16].

**Table 4.** Specific Surface Area of Nano-sized Jackfruit seed particles

| Particle Size (nm) | Surface Area (nm$^2$) | Volume (nm$^3$) | Density (g/cm$^3$) | SSA (m$^2$/g) | SA to Volume Ratio |
|---|---|---|---|---|---|
| 12 | 452 | 904 | 0.80 | 625 | 0.5 |

**3.7. XRD – Morphology Index**

It is well known that jackfruit seed powder is widely used in many diverse industries such as food, pharmaceuticals, cosmetics and paper industries. The use of jackfruit seed powder as a raw material is derived from its unique structural, physical and chemical properties, which are reflected by its hardness, surface properties, particle size and morphology. It is proposed that the specific surface area of jackfruit seed powder (which is important to many of the above mentioned industries) is dependent on the interrelationship of particle morphology and size. A XRD morphology index (MI) is developed from FWHM of XRD data to understand this relationship.

MI relates the FWHM of two peaks to its particle morphology (peak having highest FWHM and a particular peak's FWHM for which M.I. is calculated). Generally, highest FWHM peak MI is 0.5 because the MI is derived from the single peak only. MI is obtained using equation.9. Experimental jackfruit seed powder MI range is from 0.50 to 0.74 and the details are presented in Table.5. It is correlated with the particle size (range from 12 to 37nm) and specific surface area (range from 203 to 625m$^2$/g). It is observed that MI has direct relationship with particle size and an inverse relationship with specific surface area. The results are shown in figures 4 & 5.

$$M.I. = \frac{FWHM_h}{(FWHM_h + FWHM_p)} \quad \text{...............(9)}$$

Where M.I. is morphology index, $FWHM_h$ is highest FWHM value obtained from peaks and $FWHM_p$ is value of particular peak's FWHM for which M.I. is to be calculated.

**Table 5.** Morphology Index of Nano-sized Jackfruit seed particles

| FWHM (β) radians | Particle Size (D) nm | Surface Area (nm$^2$) | Volume (nm$^3$) | Specific Surface Area (m$^2$/g) | Morphology Index |
|---|---|---|---|---|---|
| 0.00930 | 15 | 707 | 1766 | 500 | 0.5565 |
| 0.00450 | 32 | 3215 | 17149 | 234 | 0.7217 |
| 0.01167 | 12 | 453 | 904 | 625 | 0.5 |
| 0.00560 | 26 | 2123 | 9194 | 288 | 0.6757 |
| 0.00405 | 37 | 4299 | 26508 | 203 | 0.7423 |

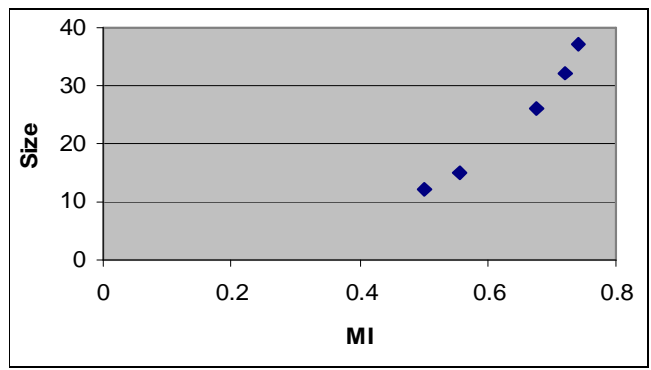

**Figure 4.** Morphological Index versus Particle Size

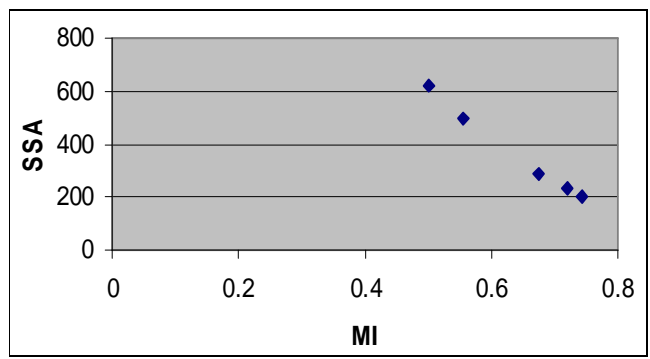

**Figure 5.** Morphological Index versus Specific Surface Area

## 4. Conclusion

Nano-sized particles of jackfruit seed has been prepared and analyzed with XRD technique successfully. XRD results conclude that jackfruit seed powder contains A- type starch particles which are semi-crystalline in nature with high specific surface area. A morphology index based on FWHM of XRD data has been developed and correlated with the particle size and specific surface area. It yields the results that MI has direct relationship with particle size and an inverse relationship with specific surface area. This study suggests that jackfruit seed powder has a lot of potential in food, cosmetics, pharmaceuticals, paper, bio-nanotechnology industries, especially its uses as thickener and binding agent. This work throws some light on and helps further research on nano-sized particles of jackfruit seed.

## Acknowledgements

The authors express immense thanks to **Dr.G.Venkadamanickam**, Rajiv Gandhi Cancer Institute & Research Center, Delhi, India, **Dr.M.Palanivelu**, *Arulmigu Kalasalingam College of Pharmacy (Kalasalingam University,* Krishnankoil, India), **S.Sivadevi,** *The SFR College for Women,* Sivakasi, India, staff & management of *PACR Polytechnic College*, Rajapalayam, India and *Ayya Nadar Janaki Ammal College*, Sivakasi, India for their valuable suggestions, assistances and encouragements during this work.